\documentclass[fleqn,aps,pra,amsmath,amssymb,preprint,showpacs]{revtex4-1}

\usepackage{float}
\usepackage{graphicx}
\usepackage{textcomp}
\usepackage{braket}
\usepackage{enumerate}

\begin{document}
\title{Permanent EDM measurement in Cs using nonlinear magneto-optic rotation}

\author{Harish Ravi}
\affiliation{Department of Physics, Indian Institute of Science, Bangalore 560012, India}

\author{Mangesh Bhattarai}
\affiliation{Department of Physics, Indian Institute of Science, Bangalore 560012, India}

\author{Abhilash Y. D. }
\affiliation{Department of Physics, Indian Institute of Science, Bangalore 560012, India}

\author{Ummal Momeen}
\altaffiliation{School of Advanced Sciences, VIT University, Vellore 632014, India}

\author{Vasant Natarajan}
\affiliation{Department of Physics, Indian Institute of Science, Bangalore 560012, India}

\begin{abstract}
We use the technique of chopped nonlinear magneto-optic rotation (NMOR) in a room temperature $^{133}$Cs vapor cell to measure the permanent electric dipole moment (EDM) in the atom. The cell has paraffin coating on the walls to increase the relaxation time. The signature of the EDM is a shift in the Larmor precession frequency which is correlated with the application of an E field. We analyze errors in the technique, and show that the main source of systematic error is the appearance of a longitudinal B field when the E field is applied. This error can be eliminated by doing measurements on the two ground hyperfine levels. Using an E field of 2.6 kV/cm, we place an upper limit on the electron EDM of $ 2.9 \times 10^{-22} $ e-cm ($95 \%$ confidence). This limit can be increased by 7 orders-of-magnitude---and brought below the current best experimental value---with easily implementable improvements to the technique. \\

\end{abstract}

\maketitle

\section{Introduction}

\begin{figure}
	\centering
    \includegraphics[width=0.9\columnwidth]{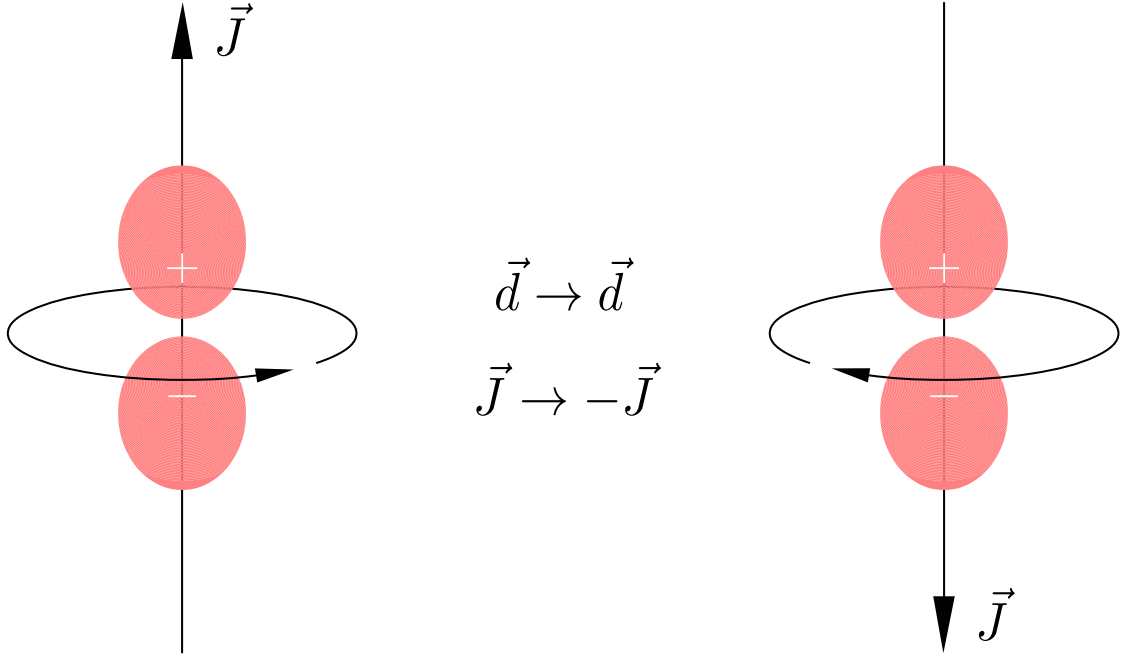}
	\caption{(Color online) Effect of time-reversal symmetry. Under T, $ \vec{d} \rightarrow d $ and $\vec{J} \rightarrow - \vec{J} $. Since the EDM vector $\vec{d}$ has to point in the direction of $\vec{J}$, it has to be zero unless T is violated.}
	\label{fig:edm_T}
\end{figure}

The search for a permanent electric dipole moment (EDM) in an atom is motivated by the fact that it signifies violation of parity (P) and time-reversal (T) symmetries in the fundamental laws of physics \cite{PUR50}. The basic idea is shown in Fig.\ \ref{fig:edm_T}. From the Wigner-Eckart theorem, we know that the total angular momentum vector is the only vector in the body-fixed frame, in the sense that all others vectors must be proportional to it. Applied to the EDM vector $\vec{d}$, this implies that
\begin{equation}
\vec{d} =d\dfrac{\vec{J}}{J}
\end{equation}
Under time reversal operation, $\vec{J}$ (which is equal to $\vec{r} \times \vec{p}$) changes sign, whereas $\vec{d}$ does not. Thus, in the time-reversed world
\begin{equation}
\vec{d} = - d \dfrac{\vec{J}}{J}
\end{equation}
which shows either that $ \vec{d} = 0 $, or that a non-zero $ \vec{d} $ implies T violation. The argument for P violation is similar---under P transformation, the (polar) vector $ \vec{d} $ changes sign, whereas the (axial) vector $\vec{J}$ does not.

Since T violation in leptons has not been observed directly \cite{foot},
EDM searches are among the most important experiments in atomic physics today. Along with charge conjugation (C), there is a theorem---the CPT theorem---which says that the combined operation of these three symmetries is always obeyed in nature \cite{STA62}. As a consequence, the observation of CP violation in neutral kaon decay \cite{CCF64} implies T violation. This is accommodated in the standard model of particle physics by predicting EDMs for fundamental particles like the electron and the neutron, but the value of the EDM predicted is about 8 to 10 orders-of-magnitude smaller than current experimental precision. However, theories that go beyond the standard model---such as supersymmetry---predict EDMs within experimental range. As a consequence, such theories are strongly constrained by experimental limits on EDMs.

The intrinsic EDM of the electron $d_e$ gets enhanced in heavy paramagnetic atoms---like Cs and Tl---due to relativistic effects \cite{SCH63,SAL64}. In fact, one of the best current limits of $|d_e| \leq 1.6 \times 10^{-27}$ e-cm has been placed through a measurement on an atomic beam of Tl atoms \cite{RCS02}. But this beam experiment is limited in precision by the systematic error arising due to the motional magnetic field $\vec{E} \times \vec{v}/c^2 $. On the other hand, polar molecules like YbF and ThO are not susceptible to this kind of error. Such molecules have the additional advantage that they are more sensitive to $d_e$ than atoms for the same value of the laboratory E field. As a consequence, experiments using the molecule YbF have resulted in a similar limit on the electron EDM of $|d_e| \leq 1.05 \times 10^{-27}$ e-cm \cite{HKS11}, and there is a proposal from this group to improve the limit using laser-cooled molecules launched in a fountain \cite{TSH13}. An order-of-magnitude improvement in precision has come from a recent measurement on the molecule ThO at the ACME collaboration, and sets a limit of $ 8.7 \times 10^{-29}$ e-cm \cite{ACME14}.

In earlier work \cite{RCN11}, we have proposed a new technique to measure the existence of an EDM in Cs atoms using a paraffin-coated vapor cell. This method is also not susceptible to errors due to the motional magnetic field because the \textbf{average} velocity inside a vapor cell is zero. Our technique is called chopped nonlinear magneto-optic rotation (chopped-NMOR), and allows us to measure the Larmor precession frequency with high precision. The signature of an EDM is a shift in the Larmor frequency correlated with the application of a large E field. 

In this work, we demonstrate the use of this technique to do an EDM measurement with an applied E field of 2.6 kV/cm. From this, we put an upper limit on the atomic EDM of $ 3.5 \times 10^{-20}$ e-cm ($95 \%$ confidence), which implies that $|d_e| \leq 2.9 \times 10^{-22}$ e-cm using the enhancement factor of 120.53 in Cs \cite{NSD08}. Though this is not competitive with the best measurement for the electron EDM, there are foreseeable improvements to the technique which will enable us to reach a precision better than the current best value.

\section{Theoretical Analysis}
As shown in our earlier work in Ref.\ \cite{RCN11}, the chopped NMOR technique relies on the laser beam measuring the rotation being modulated on and off. During the on time, the atoms are optically pumped into a $\Delta m =2 $ coherence of the ground state (atomic alignment). During the off time, the atoms freely precess around the magnetic field at the Larmor frequency $\omega_L$. If the chopping frequency matches $2\omega_L$ (the factor of 2 appears because the alignment has 2-fold symmetry), then the rotation is resonantly enhanced in every cycle.

Applying the Wigner-Eckart theorem to the total angular momentum $\vec{F}$ of the atom, both the magnetic moment $\vec{\mu}$ and electric dipole moment $\vec{d}$ have to point along $\vec{F}$. Hence the interaction energy of the atom in the presence of the two fields is
\begin{equation}
U = \left(d\vec{E} + \mu \vec{B}\right) \cdot \frac{\vec{F}}{F}
\end{equation}
The quantization axis of the atoms is defined by the direction of the B field. The Larmor precession frequency is determined by the energy spacing between two adjacent magnetic sub levels, i.e.\ $ \Delta m_F = 1 $. It is therefore given by 
\begin{equation}
\omega_L = (g_F \mu_B B)/\hbar
\end{equation}
where $g_F$ is the Land\'e $g$ factor of the level and $\mu_B$ is the Bohr magneton. In the presence of an additional E field---in the same direction to maintain the quantization axis---the Larmor precession frequency changes to 
\begin{equation}
\omega_L = (g_F \mu_B B + d_e \eta E)/\hbar
\end{equation}
where $d_e$ is the permanent EDM of the electron and $\eta $ is the EDM enhancement factor in the atom. This shows that if the atom has an electric dipole moment, then there will be a change in the Larmor precession frequency when the E field is applied. 

In a chopped-NMOR experiment, the chopping frequency is kept constant while the B field is scanned. In the absence of an E field, we get two peaks at $B_+$ and $B_-$, such that 
\begin{equation}
B_+ = \dfrac{\hbar \omega_L}{2g_F\mu_B}
\end{equation} 
\begin{equation}
B_- = \dfrac{\hbar \omega_L}{-2g_F \mu_B} = - B_+
\end{equation}
In the presence of an additional E field, and since the chopping frequency is the same, the peaks move to a new location given by 
\begin{equation}
\begin{aligned}
\hbar \omega_L &= 2g_F \mu_B B_+^E + 2d_e \eta E  \\
\implies B_+^E &= \dfrac{\hbar \omega_L - 2d_e \eta E}{2g_F \mu_B} 
\end{aligned}
\end{equation}
\begin{equation}
\begin{aligned}
\hbar \omega_L &= \left| -2g_F \mu_B B_-^E + 2d_e \eta E \right| \\
\implies B_-^E &= \dfrac{\hbar \omega_L + 2d_e \eta E}{2g_F \mu_B}
\end{aligned}
\end{equation}
using the fact that the E term is much smaller than the B term. 

One way to understand this is that the quantization axis---set by the B field---is opposite for the two signs of the B field, therefore, the constant E field contributes with different signs to the two values.

\subsection{Separation and center of peaks}

From the above expressions, the separation of the peaks are
\begin{equation}
\begin{aligned}
S_{\rm no \, E} &= B_+ - B_- = \frac{\hbar \omega_L}{\mu_B g_F} \\
S_{\rm with \, E} &= B_+^E - B_-^E = \frac{\hbar \omega_L}{\mu_B g_F} = S_{\rm no \, E}
\end{aligned}
\end{equation}
and the corresponding centers are 
\begin{equation}
\begin{aligned}
C_{\rm no \, E} &= \dfrac{1}{2} \left[B_+ + B_-\right] = 0  \\
C_{\rm with \, E} &= \dfrac{1}{2} \left[ B_+^E + B_-^E \right] = - \dfrac{2d_e\eta E}{g_F \mu_B}
\end{aligned}
\end{equation}
Thus the EDM signal appears as a change in the center of the two peaks proportional to the E field, with no concomitant change in the peak separation.

\subsection{Effect of stray B fields}
In order to understand this effect, we assume that there is a stray magnetic field with a longitudinal component $B_{\ell}$ and a transverse component $B_t$. Since the atom precesses around the magnitude of the total field, the Larmor precession frequency changes. For a given chopping frequency, one gets new peaks with a separation of
\begin{equation}
S^{\rm stray} = 2\sqrt{\left(\dfrac{\hbar \omega_L}{2g_F \mu_B}\right)^2 - B_t^2}
\end{equation}
and a center given by
\begin{equation}
C^{\rm stray} = -B_{\ell}
\end{equation}
Thus, a stray magnetic field will cause both the separation and center of the peaks to change, but only the longitudinal component will mimic the EDM signal.

\section{Experimental details}

\begin{figure*}
\centering
\includegraphics[width=0.8\textwidth]{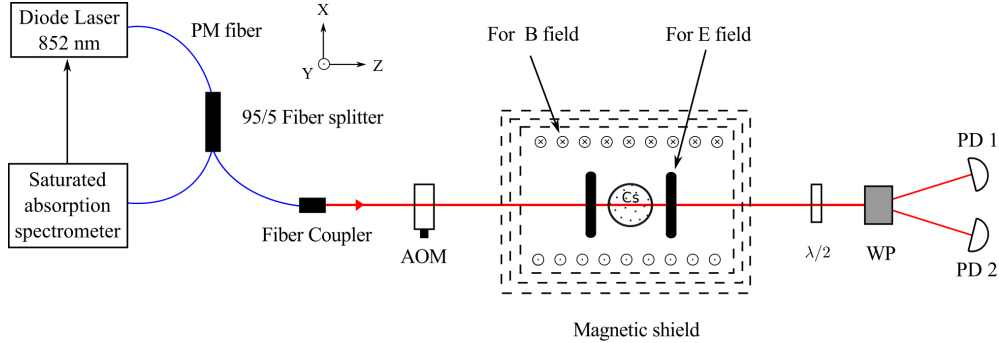}
\caption{(Color online) Schematic of the experiment. Figure key: PM fiber -- polarization maintaining fiber; AOM -- acousto-optic modulator; $\lambda/2 $ -- half-wave retardation plate; WP -- Wollaston prism; PD -- photodiode.}
\label{fig:edm_setup}
\end{figure*}

The experimental setup is shown schematically in Fig.\ \ref{fig:edm_setup}. The laser beam is derived from a commercial diode laser system (Toptica DL Pro), operating on the 852 nm D$_2$ line of Cs (6S$_{1/2} \rightarrow $ 6P$_{3/2}$ transition). The isotope used for the experiment has a mass number $M = 133$, and nuclear spin $I = 7/2$. The output of the laser comes out of a polarization maintaining fiber. The fiber goes into a 95/5 power splitter, with 5\% of the power fed into a compact saturated-absorption spectroscopy (CoSy) unit. The signal from the unit coupled with current modulation of the diode laser produces an error signal so that the laser can be locked to any peak in the CoSy spectrum. For the experiment, the laser is locked to the $ 4 \rightarrow (4,5) $ crossover resonance of $^{133}$Cs, which is about 125.5 MHz below the $ 4 \rightarrow 5 $ hyperfine transition. This frequency shift is provided by an acousto-optic modulator (AOM). The AOM serves two additional purposes: (i) providing control of the beam power by controlling the RF power going into the AOM, and (ii) allowing chopping of the laser beam by doing on-off modulation of the AOM driver. The frequency of the AOM is set using a function generator (HP 8656B). The timebase of the function generator is set using a commercial Rb clock (SIM 940) with $ 10^{-12} $ relative stability. The frequency output of the function generator is internally modulated for the chopping, and its output amplitude is computer controlled to adjust the beam power.

The remaining 95\% of the laser output (after the power splitter) is coupled to free space using a fiber coupler. The beam size after the coupler ($1/e^2$ diameter) is 3 mm. The RF power driving the AOM is adjusted to get a laser beam power of 100 \textmu W. It goes into a spherical Cs vapor cell with 75 mm diameter and paraffin coating on the walls. The cell is inside a 3-layer magnetic shield (Magnetic Shield Corp, USA) with a shielding factor of better than $ 10^4 $. The longitudinal B field required for the chopped NMOR measurement is achieved by placing the cell inside a solenoid coil. The coil is wound on an acrylic form of 190 mm diameter, and consists of 1800 turns of 0.35 mm magnetic wire wound over a length of 640 mm. The voltage on the solenoid coil is set using a 24 bit NI DAQ card with a range of $-3.5$ to $ +3.5 $ V. This voltage is varied to scan the B field.

The plane of polarization of linearly polarized light gets rotated in the vapor cell due to the phenomenon of NMOR. The output beam is split into its two polarization components using a Wollaston prism. The power in each component is measured using photodiodes. The powers in the two components are made equal (in the absence of an NMOR signal) by adjusting the $ \lambda/2$ waveplate after the cell. When the angle of rotation is small, the optical rotation can be shown to be proportional to the difference in the two PD signals \cite{BGK02}. Hence, the difference signal is demodulated in a commercial lock-in amplifier (SR 830). The out-of-phase component of the lock-in amplifier is used in the experiment. The data is acquired to the computer using the same DAQ card used for setting the solenoid voltage.

The E field required for the EDM measurement is applied using two plates with equal and opposite high voltage on them. This ensures that the value of the high voltage at each plate is only half of what is required for a given E field, and hence ensures that breakdown at each plate is minimized. The plates are made of polished aluminium and spaced 85 mm apart using teflon rods. They have dimensions of 130 mm diameter and 40 mm thickness, with rounded edges to minimize the E field magnitude at the corners. The plates also have a 6 mm hole at their centers to allow the laser beam to pass through. 

\section{Error analysis}

\subsection{Statistical errors}
Statistical errors arise due to the following.
\begin{enumerate}[(i)]
	\item Errors in the curve-fitting routine used for determining the peak centers, arising due to the finite linewidth and finite signal-to-noise ratio (SNR) of the peaks.
	
	\item Random variation in the location of the peaks due to jitter in the function generator used for driving the AOM. This is minimized by locking the function generator to an ultra-stable Rb atomic clock, as described in the ``Experimental details'' section.
	
	\item Errors in the voltage applied to the solenoid coil used for scanning the B field.
	
	\item Shot-to-shot variation in the stray B field in the lab, arising due to movement of heavy metallic objects. Note that a constant B field will not cause an error because the effect will be the same from one measurement to the next. This is important because there are many sources of stray B fields in our building, but fortunately most of them are static. 
\end{enumerate}
The good thing about statistical errors is that they average to zero. Thus, all the above sources of error can be reduced below any required level by taking more data. But this makes sense only if systematic errors can be kept below the same level.

\subsection{Systematic errors}
The analysis in the ``Theoretical analysis'' section shows that the main source of systematic error will be the appearance of a \textbf{longitudinal} B field correlated with  the application of the E field, because other errors such as due to E and B field gradients are negligible. One potential source of such an E-correlated field is leakage current between the plates when the high voltage is applied. Note that a leakage current along the rods holding the two plates will only cause a transverse field, and hence will not affect the EDM measurement. However, there could be a small current that makes it through the air gap between each plate and the solenoid coil. This will cause a longitudinal B field, and hence mimic the EDM signal. 

However we have an experimental handle to deal with this error, namely to repeat the experiment with the other $F$ level of the ground state of $^{133}$Cs, namely $F_g = 3 $. The cancellation of systematic error due to a longitudinal B field relies on the fact that the Land\'e $ g $ factor values for the two levels are equal and opposite. From our previous analysis, the shift in the peak center in the presence of both E and B fields is 
\begin{equation}
\Delta C = - \dfrac{2 d_e \eta E}{g_F \mu_B} - B_\ell
\end{equation}
This shows that the shift due to the E field will be opposite for the two ground levels (since they have $ g_F $ values of $+1/4$ and $-1/4$), but the shift due to the B field will be the same. Thus the difference between the two measurements will give us an EDM measurement without the systematic error due to the presence of a time-varying B field. In effect, the other level acts as a co-magnetometer, similar to the use of Rb atoms in a Cs vapor cell as demonstrated by us in Ref.\ \cite{RCN11} and proposed by others for EDM measurements \cite{CLV01}. The use of the other hyperfine level is a novel feature of our technique, and is reported for the first time for atomic EDM measurements.

\section{Experimental results}

\begin{figure}
\centering
\includegraphics[width=.9\columnwidth]{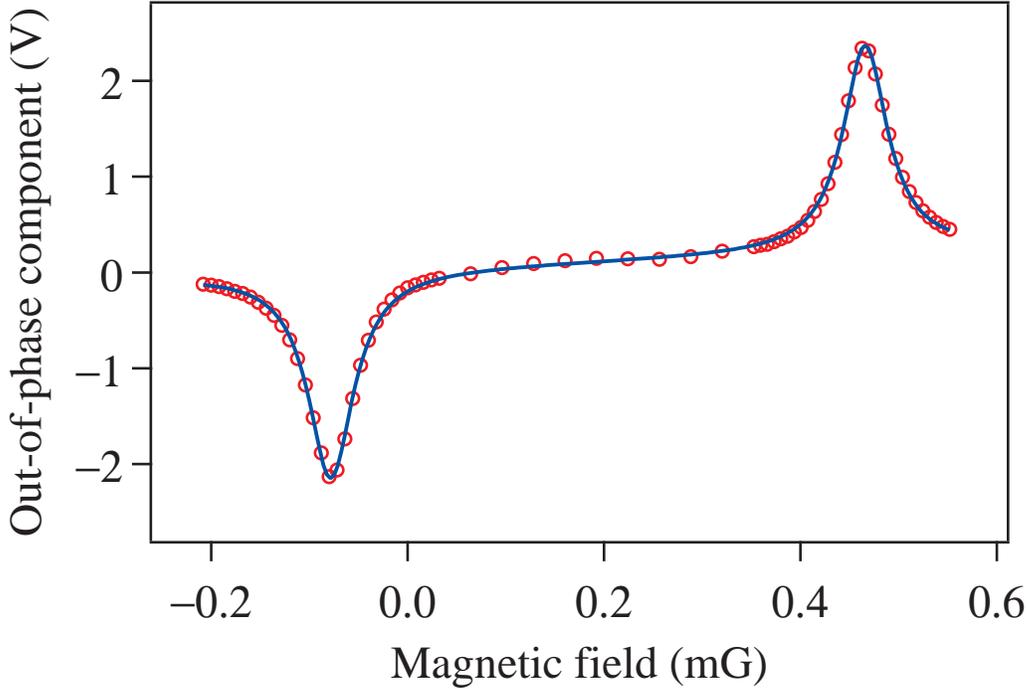}
\caption{(Color online) Output of the lock-in amplifier showing the out-of-phase component of the difference signal, obtained at a chopping frequency of 195 Hz. Two well-resolved Lorentzian peaks appear as the B field is scanned---one at negative field and one at positive field. The solid line is a curve fit to two Lorentzian peaks along with a linear baseline (to account for the linear MOR effect), which yields the peak centers.}
\label{fig:nmor}
\end{figure}

The experiment consists of measuring multiple curves of the kind shown in Fig.\ \ref{fig:nmor}. For each curve, the voltage on the solenoid coil is varied from $ -2 $ to $ +3.5 $ V in variable steps. The slight asymmetry in the two peaks is because of imperfect shielding by the magnetic shield, which results in a small residual longitudinal B field. This is also evident in the curve shown from the fact that the center of the two peaks is located not at 0 but at the point where the applied field cancels the residual field. At each value of the voltage, the output of the lock-in amplifier is averaged 1000 times at intervals of 1 ms, therefore the measurement at each point lasts 1 s. Fig.\ \ref{fig:nmor} shows that the density of points along the $ x $ axis is not constant. This is because there are more points near the peaks where there is more signal, than near the zero crossing where there is negligible signal. The total number of points in each curve is 70, therefore it takes 70 s to complete a curve. The solid line in the figure is a curve fit to two Lorentzian peaks along with a linear baseline (to account for the linear MOR effect). The curve fit yields the centers of the two Lorentzian peaks, from which the curve center is determined. This process is repeated 18 times. The average value of this set then yields a value of $ C_{\rm no \, E} $. The same procedure is repeated with an E field applied, which gives a value of $ C_{\rm with \, E} $. Thus, each set consists of 18 values of $ C_{\rm no \, E} $, and 18 values of $ C_{\rm with \, E} $. The difference $ \Delta C = C_{\rm with \, E} - C_{\rm no \, E} $ combined with the value of the E field then yields one measurement of the EDM as follows
\begin{equation}
d_e = \dfrac{g_F \mu_B}{2 \eta E} \, \Delta C
\end{equation}
The relevant values for our case are $ g_F = +1/4 $, $ \eta = 120.53$, and $ E = 2.6 $ kV/cm.

The first set of measurements comprised of measurements of $ d_e $ a total of 50 times. The average value and standard deviation---statistical error---yields
\begin{equation*}
d_e = (0.3 \pm 1.3) \times 10^{-22} \text{ e-cm}
\end{equation*}

The main source of systematic error in our measurement is the appearance of a leakage current through the air gap between the plates and the solenoid. Since the leakage current appears only when the high voltage is applied, it is correlated with the application of the E field. However, only that part of the leakage current that makes it through the solenoid will appear as correlated longitudinal B field, and hence mimic the EDM signal. Leakage current through the teflon rods holding the two plates, and any charging currents during the transient build up of the E field, will only contribute to a transverse field, and hence not appear as a systematic error.

We estimate the size of this error by doing the following calculation. Assuming for simplicity that the plate is a cylinder of finite length placed within another cylinder of infinite length, the resistance of the air gap is 
\begin{equation}
R_{\rm air} = \dfrac{\rho}{2 \pi} \dfrac{\ln \left( d_2/d_1 \right)}{t}
\end{equation}
where $\rho$ is the resistivity of air, $d_2$ is the inside diameter of the solenoid, $d_1$ is the outside diameter of each field plate, and $t$ is the thickness of the plate. For our dimensions, the value of resistance is $ R_{\rm air} = 4.5 \times 10^{16} $ \textohm, which causes a leakage current of the order of $ 0.5 $ pA. This will cause a leakage field of 
\begin{equation*}
B_{\rm leak} = 1.8 \times 10^{-11} \text{ G}
\end{equation*}
which will result in an EDM of 
\begin{equation*}
d_e^{\rm syst} = \dfrac{g_F \mu_B B_{\rm leak}}{2 \eta E} \approx 4.2 \times 10^{-26} \text{ e-cm}
\end{equation*}
Since this is 4 orders of magnitude smaller than the statistical error in our measurement, we use only the statistical error in the EDM measurement. From this, we put an upper limit on the electron EDM of 
\begin{equation*}
\lvert d_e \rvert \leq 2.9 \times 10^{-22} \text{ e-cm}
\end{equation*}
The standard deviation has been increased by a factor of 1.96 to get a $ 95 \% $ confidence level.

\section{Future improvements}

The following improvements in the technique will enable us to increase the precision of the measurement.
\begin{enumerate}[(i)]
\item Better shield -- the current 3-layer shield has a shielding factor of $ 10^4 $. There is a 4-layer design described in Ref.\ \cite{BGK02}, which will improve the shielding by a factor of 100. This should give us a corresponding decrease in the linewidth, and hence reduce the uncertainty in determining peak centers.

\item Better cell -- The present cell has paraffin coating on the walls. We next plan to use a cell with alkene coating on the walls, as described in Ref.\ \cite{BKL10}. This will increase the relaxation time by a factor of 100, and result in a correspondingly smaller linewidth.

\item Higher E field -- The current field is limited by electrical breakdown of air at sharp points on the field plates. We expect to use a field that is 100 times larger by evacuating the entire assembly. The increased field will result in a direct improvement in the precision of the measurement. This will also result in a smaller leakage current.

\item Longer time -- As mentioned before, each curve takes about a minute to complete. Therefore, the entire set of 1800 curves is completed in less than 1 day. This can easily be increased by a factor of 100, which should result in a increase in statistical precision of a factor of 10.
\end{enumerate}
Together, the above improvements will result in a factor of $ 10^7 $ increase in \textit{statistical} precision. If systematic errors can also be kept under control---maybe by measuring on the other ground hyperfine level---then our measurement has the potential to become the best EDM measurement.

\section{Conclusions}

In summary, we have demonstrated a new technique for the measurement of the electron EDM. The method uses chopped NMOR in an anti-relaxation coated Cs vapor cell at room temperature. We analyze errors in the technique, and show that a leading source of systematic error (due to an E-correlated longitudinal B field) can be eliminated by doing measurements on the two ground hyperfine levels. One significant advantage of our technique compared to other methods is that the charging current as the E field builds up will cause only a transverse B field to appear, which is not a source of systematic error.

Using an E field of 2.6 kV/cm, we set a limit of $\lvert d_e \lvert \leq 2.9 \times 10^{-22}$ e-cm. The most precise limit comes from a measurement on the molecule ThO, which is about 6 orders-of-magnitude below our limit. However, there are easily implementable improvements to the technique that should enable us to reach a precision below this value.

\section*{Acknowledgments}
This work was supported by the Department of Science and Technology, India. The authors thank S Raghuveer for help with the manuscript preparation.


\begin{thebibliography}{15}
\bibitem{foot}
T violation in hadrons has been observed directly in the decay of B mesons in the BaBar experiment.
\end{thebibliography}
\end{document}